\documentclass[a4paper,11pt]{article}
\pdfoutput=1 

\usepackage{jheppub} 
\usepackage[T1]{fontenc} 
\usepackage{physics}

\title{A Generalized Uncertainty Principle from a Mediating Field}

\author[a,1]{Dor Gabay\note{Corresponding author.}}

\affiliation[a]{Department of Physical Electronics, Tel-Aviv University, Tel-Aviv 69978, Israel.}

\emailAdd{dorgabay@mail.tau.ac.il}

\abstract{A generalized uncertainty principle is obtained from a conformally transformed action containing a scalar field and a unique constraint. 
The constraint's Lagrange multiplier is found to obey a relativistic diffusion equation transforming the internal coordinates of the scalar field, via the shift theorem. 
For an approximately conserved Noether current, the coupled wave- and diffusion-like equations are merged into an infinite-order partial differential equation (PDE). 
It is conjectured that infrared and ultraviolet divergences are naturally removed in studying the infinite-order PDE's corresponding propagator.
It is further suggested that higher-order contributions of the associated commutation relations are reminiscent of a self-interaction not present in the current quantum theory.}

\begin{document} 
\maketitle
\flushbottom

\section{Introduction}
Heisenberg's uncertainty principle has been the underlying architecture of quantum mechanics for nearly a century~\cite{robertson1929UP}. 
In the framework of quantum field theory, it has proved to be so reliable, that no other principle has led to such an order of accuracy of the fundamental constants~\cite{mohr2016constants}. 
Although, once a system is imbursed in immensely high energies, gravitational fields dwelling at correspondingly smaller length scales begin to hinder at the need for an alternate theory~\cite{rovelli2008lqg,nicolai2005lqg,aharony2000string}. 
It therefore suffices to presume that a natural extension of the uncertainty principle should exist, one which incorporates gravitational phenomena at higher energies and converges to Heisenberg's uncertainty principle at lower energies. 
Various extensions of the uncertainty principle have been proposed by superstring theories~\cite{adler1999UP,tawfik2015UP}, whereby Planck scale corrections are considered in accounting for the gravitational interaction of the electron with the photon. 

In the work which follows, we suggest that a certain class of conformally transformed actions can naturally serve as extensions to Heisenberg's uncertainty principle. 
Conformal theories have attracted much attention in recent years~\cite{francesco2012cft,gaberdiel2000cft,WeylReview2017}. 
Within the framework of field theory~\cite{francesco2012cft,gaberdiel2000cft}, they have successfully been used to characterize two dimensional critical points and closed strings constructed by oppositely propagating waves composed of chiral fields. 
Within the framework of gravity (e.g. Weyl theory~\cite{WeylReview2017}), they naturally incorporate the early accelerated expansion of space-time. 
Conformal theories of gravity have also been shown to contain parity and time reversal symmetry~\cite{Bender1,Bender2,Manheim_PTSym}, resulting in a broader class of Hamiltonians beyond that of Hermiticity and hindering at the possible advantages of geometrical interpretations of quantum mechanics~\cite{Sidharth_QM,QRGM_Susskind,QMGeometry}. 

One such geometric interpretation of quantum mechanics is that of Bohm~\cite{BohmI,BohmII,PHollandBook,SheldonReview}. The Bohmian theory of quantum mechanics disassociates a scalar field's density and phase to interpret its space-time trajectories. 
Within the framework of conformal gravity, the Bohmian interpretation has been demonstrated to inherit a unique correspondence to the metric tensor, via the Hamilton-Jacobi equation~\cite{our_cosmo_paper,Shojai_Article,ShojaiBohmianQM}. 
In considering the non-conserved properties of such a conformally transformed action~\cite{our_cosmo_paper}, quantum mechanics can be extended to include diffusive properties resulting in a generalized uncertainty principle. 

To demonstrate this, coupled wave- and diffusion-like equations of motion are first attained in varying a conformally transformed Bohmian action~\cite{our_cosmo_paper}.
The wave-like equation resembles that of Klein-Gordon with an added source and dissipation term.
The dissipation term takes the form of a non-conserved Noether current while the source term arises from an added constraint within the action, used in bypassing Weinberg's no-go theorem~\cite{WeinbegCosmo89}. 
The constraint's Lagrange multiplier, hereon referred to as the mediating field, is suggested to mediate the source and dissipation terms within the scalar field equation. 
The mediating field's diffusion-like equation of motion is further shown to transform the internal coordinates of the scalar field. 
When the conservation conditions for the scalar and mediating fields are approximately satisfied, the coupled wave- and diffusion-like equations of motion merge into an infinite-order PDE. 
Canonical commutations relations containing higher-order contributions are then heuristically derived and, following that, a generalized uncertainty principle. 
A representation theorem is proposed, suggesting that the corrections to the scalar field can either be articulated by an extended momentum operator or differential form. 
To conclude, in the limit of conserved scalar and mediating fields, a propagator is derived and conjectured to alleviate infrared (IR) and ultraviolet (UV) divergences.

\section{Theory}
In previous works~\cite{our_cosmo_paper}, an Einstein-Hilbert action, composed of a Klein-Gordon field and a unique constraint, was conformally transformed in order to bypass Weinberg's no-go theorem~\cite{WeinbegCosmo89}. The constraint's Lagrange multiplier $\lambda$, hereon referred to as the mediating field, ensures stationary solutions of the tensor and scalar equations of motion. Written in terms of the Einstein-Hilbert Lagrangian $\mathcal{L}_{EH}$, metric tensor $g_{\mu\nu}$, conformal factor $\Omega^{2}$, and scalar field $\varphi$, the action takes the form
\begin{eqnarray}
A[g_{\mu\nu},\Omega,\varphi,\lambda]=&&\frac{1}{2\kappa}\int\mathrm{d}^{4}x\sqrt{-g}\mathcal{L}_{EH}(g_{\mu\nu},\Omega^{2})+\Upsilon_{0}\int\mathrm{d}^{4}x\sqrt{-g}\Omega^{2}\Bigl[\nabla_{\mu}\varphi^{*}\nabla^{\mu}\varphi-\frac{m^{2}}{\hbar^{2}}\Omega^{2}\varphi^{*}\varphi\Bigr] \nonumber \\
&&-\int\mathrm{d}^{4}x\sqrt{-g}\lambda\Bigl[\alpha_{0}\mathrm{ln}\Omega^{2}-\mathcal{W}(\varphi,\varphi^{*},\nabla_{\mu}\varphi,\nabla_{\mu}\varphi^{*},...)\Bigr] \label{vacuaA}
\end{eqnarray}
Here, and in the remainder of this work, the speed of light $c=1$ for convenience. Unlike Weyl-squared theories~\cite{boulanger2001weyl}, the conformally transformed Einstein-Hilbert action is not conformally invariant, and is hereon studied in its non-conformal frame of reference (i.e. conformal factor is disassociated from metric tensor). The conformal factor $\Omega^{2}$ is fixed to $\mathcal{W}$, a function which must obey a certain set of conditions~\cite{our_cosmo_paper} to satisfactorily bypass Weinberg's no-go theorem; one such condition being that the constant $\alpha_{0}$ must be specified so as to ensure that the action remains dimensionless. Additionally, the unknown constant $\Upsilon_{0}$ is used to properly scale the Klein-Gordon action with the newly considered constraint. 

By setting the function $\mathcal{W}$ to the pre-factor of the effective mass of the relativistic Hamilton-Jacobi equation, the conformal factor is enforced to conform with a density-dependent relativistic quantum potential $Q=\ell_{c}^{2}\Box\sqrt{\rho}/\sqrt{\rho}$
\begin{eqnarray}
\Omega^{2}=\textup{e}^{Q}\;\;\to\;\;\mathcal{W}(\sqrt{\rho},\nabla_{\mu}\sqrt{\rho},\Box\sqrt{\rho},...)=\frac{\Box\sqrt{\rho}}{\sqrt{\rho}}. \label{constraintQ}
\end{eqnarray}
Here, $\rho=\varphi^{*}\varphi$ is the particle density and $\alpha_{0}$ is set to the inverse Compton wavelength squared $\ell_{c}^{-2}=\frac{m^{2}}{\hbar^{2}}$ to ensure a dimensionless action. Given the resulting square bracketed term of the constraint has a dimensionality of two, the dimensionality of $\lambda$ must be two in a four dimensional space-time, analogous to that of a density. To more elegantly vary the density-dependent functional, we choose to dwell within the Bohmian framework of quantum mechanics~\cite{BohmI,BohmII}, resulting in an expression analogous to (\ref{vacuaA})
\begin{eqnarray}
A[g_{\mu\nu},{\Omega}, S, \rho, \lambda]&=&
\frac{1}{2\kappa}\int{d^4x\sqrt{-g}\mathcal{L}_{EH}(g_{\mu\nu},\Omega^{2})}+\Upsilon_{0}\int{d^4x\sqrt{-g}\frac{1}{\hbar^{2}}\left[\Omega^{2}\rho\nabla_{\mu}S\nabla^{\mu}S-m^{2}\Omega^4\rho\right]} \nonumber \nonumber \\
&&-\int{d^4x\sqrt{-g}\lambda\left[\frac{m^{2}}{\hbar^2}\ln{\Omega^2}-\frac{\Box\sqrt{\rho}}{\sqrt{\rho}}\right]}. \label{Actioneq}
\end{eqnarray}
Here, $S$ is Hamilton's principle function, hereon referred to as the `phase' for simplicity. In varying the above action, it is important to pay careful attention to both the volumetric and boundary terms. While the volumetric terms characterize the equations of motion, the boundary terms play a crucial role in identifying the underlying conservation conditions in attaining the equations of motion. In varying the Lagrangian $\mathcal{L}$ with respect to both $\sqrt{\rho}$ and $S$, one arrives at the following prescriptions 
\begin{eqnarray}
\frac{\delta\mathcal{L}}{\delta{\sqrt{\rho}}}=\Biggl[\frac{\partial\mathcal{L}}{\partial{\sqrt{\rho}}}+\Box\Bigl(\frac{\partial\mathcal{L}}{\partial{\Box\sqrt{\rho}}}\Bigr)\Biggr]\delta\sqrt{\rho}-\Biggl[2\nabla_{\mu}\Bigl(\nabla^{\mu}\Bigl(\frac{\partial\mathcal{L}}{\partial{\Box\sqrt{\rho}}}\Bigr)\delta\sqrt{\rho}\Bigr)-\Box\Bigl(\frac{\partial\mathcal{L}}{\partial{\Box\sqrt{\rho}}}\delta\sqrt{\rho}\Bigr)\Biggr] \label{varrho}
\end{eqnarray}
\begin{eqnarray}
\frac{\delta\mathcal{L}}{\delta{S}}=\Biggl[\frac{\partial\mathcal{L}}{\partial{S}}-\nabla^{\mu}\Bigl(\frac{\partial\mathcal{L}}{\partial{\nabla^{\mu}S}}\Bigr)\Biggr]\delta{S}+\Biggl[\nabla^{\mu}\Bigl(\frac{\partial\mathcal{L}}{\partial{\nabla^{\mu}{S}}}\delta{S}\Bigr)\Biggr]. \label{varS}
\end{eqnarray}
The equation of motion for the density and phase resulting from the left square brackets of (\ref{varrho}) and (\ref{varS}), respectively, take the form
\begin{eqnarray}
\nabla_{\mu}S\nabla^{\mu}S-m^2\Omega^2+\frac{\hbar^{2}\Upsilon_{0}^{-1}}{2\Omega^2\sqrt{\rho}}\Bigl[\Box{\Bigl(\frac{\lambda}{\sqrt{\rho}}\Bigr)}-\lambda\frac{\Box\sqrt{\rho}}{\rho}\Bigr]=0 \label{EoM}
\end{eqnarray}
\begin{eqnarray}
\nabla_{\mu}(\Omega^2\rho\nabla^{\mu}S)=0. \label{cnteqn}
\end{eqnarray}
Here, (\ref{cnteqn}) is simply the Noether current. The terms in the right square brackets of (\ref{varrho}) and (\ref{varS}) predetermine the conservation conditions of the theory. The conservation condition of (\ref{varS}) is not unique as it results in the same expression as that of (\ref{cnteqn}), but (\ref{varrho}) can be quite insightful
\begin{eqnarray}
\lim\limits_{\delta\sqrt{\rho}\to\sqrt{\rho}}\Biggl[2\nabla_{\mu}\Bigl(\nabla^{\mu}\Bigl(\frac{\partial\mathcal{L}}{\partial{\Box\sqrt{\rho}}}\Bigr)\delta\sqrt{\rho}\Bigr)-\Box\Bigl(\frac{\partial\mathcal{L}}{\partial{\Box\sqrt{\rho}}}\delta\sqrt{\rho}\Bigr)\Biggr]=0 \label{cnstrnt0}
\end{eqnarray}
\begin{eqnarray}
\Box\lambda=2\nabla_{\mu}\Big(\lambda\frac{\nabla^{\mu}\sqrt{\rho}}{\sqrt{\rho}}\Big). \label{cnstrnt1}
\end{eqnarray}
The resulting expression suggests that the conservation condition is only satisfied in the limit of $\lambda\to\rho$. This condition is also evident in (\ref{EoM}), whereby the square bracketed term nullifies in the limit of $\lambda\to\rho$. The alignment of the mediating field with the density is a conservation condition heron referred to as the \textit{field alignment scenario}. Whether this condition is satisfied in every scenario is not immediately apparent, but suggests that (\ref{cnstrnt0}) should approximately hold even when $\lambda$ and $\rho$ are not perfectly aligned. 

The tensor properties of the mediating field are just as pertinent. In varying the action with respect to the metric tensor and conformal factor, tracing the former, and equating the scalar curvatures of the two, one arrives at an equation of motion for the mediating field $\lambda$
\begin{eqnarray}
\nabla_{\mu}\Bigl(\lambda\frac{\nabla^{\mu}\sqrt{\rho}}{\sqrt{\rho}}\Bigr)=\frac{m^2}{\hbar^2}(1-Q)\lambda. \label{LambdaEq}
\end{eqnarray}
In instead adopting a linear-order conformal factor $\Omega^{2}=(1+Q)$ within the constraint of (\ref{Actioneq}), the equation of motion for the mediating field takes a surprisingly similar form
\begin{eqnarray}
\nabla_{\mu}\Bigl(\lambda\frac{\nabla^{\mu}\sqrt{\rho}}{\sqrt{\rho}}\Bigr)=\frac{m^2}{\hbar^2}\lambda. \label{LambdaEqlin}
\end{eqnarray}
To no surprise, the conservation condition corresponding to each of the constraints are the same. This expression will seldom be used throughout the text to identify elements of the theory which would otherwise be difficult to recognize. 
It is the prior goal of this manuscript to explore the underlying meaning of these coupled equations of motion in their scalar field form. In the sections which proceed, we will demonstrate that (\ref{LambdaEq})-(\ref{LambdaEqlin}) result in diffusion equations mediating the source and dissipation terms contained within the wave-like equation of the scalar field.

\section{Wave Equation}
Whether equations of motion are studied within their Bohmian or scalar field form should not significantly matter. 
Typically, the scalar field form is preferred due to its inclusion of the continuity equation within a single equation of motion.
In what follows, the Bohmian equations (\ref{EoM})-(\ref{cnteqn}) are re-expressed into their scalar field form $\varphi=\sqrt{\rho}\textup{e}^{iS/\hbar}$. To begin with, we assume there to be no constraint within the action, reducing the equation of motion for the density to
\begin{eqnarray}
\nabla_{\mu}S\nabla^{\mu}S=m^2\Omega^2. \label{EoMsimp}
\end{eqnarray}
Given the corresponding identity 
\begin{eqnarray}
\nabla^{\mu}S = -i\hbar\Bigl(\frac{\nabla^{\mu}\varphi}{\varphi}-\frac{\nabla^{\mu}\sqrt{\rho}}{\sqrt{\rho}}\Bigr), \label{Sexprsn}
\end{eqnarray}
the continuity equation (\ref{cnteqn}) can be re-expressed into a the following relation
\begin{eqnarray}
\frac{i}{\hbar}\nabla_{\mu}(\rho\Omega^2\nabla^{\mu}S)=\nabla_{\mu}\Bigl(\Omega^{2}\rho\frac{\nabla^{\mu}\varphi}{\varphi}\Bigr)-\nabla_{\mu}\Bigl(\Omega^{2}\rho\frac{\nabla^{\mu}\sqrt{\rho}}{\sqrt{\rho}}\Bigr)=0 \label{cnt_idnty0}
\end{eqnarray}
\begin{eqnarray}
\frac{\nabla_{\mu}\varphi\nabla^{\mu}\varphi}{\varphi^{2}}+\frac{\nabla_{\mu}\sqrt{\rho}\nabla^{\mu}\sqrt{\rho}}{\sqrt{\rho}^{2}}-2\frac{\nabla_{\mu}\sqrt{\rho}}{\sqrt{\rho}}\frac{\nabla^{\mu}\varphi}{\varphi}=\frac{\Box\varphi}{\varphi}-\frac{\Box\sqrt{\rho}}{\sqrt{\rho}}+\frac{i}{\hbar}\frac{\nabla_{\mu}\Omega^{2}}{\Omega^{2}}\nabla^{\mu}S. \label{cnt_idnty1}
\end{eqnarray}
In then expanding the equation of motion of the density, one gets
\begin{eqnarray}
\nabla_{\mu}S\nabla^{\mu}S=-\hbar^{2}\Bigl(\frac{\nabla^{\mu}\varphi}{\varphi}-\frac{\nabla^{\mu}\sqrt{\rho}}{\sqrt{\rho}}\Bigr)\Bigl(\frac{\nabla^{\mu}\varphi}{\varphi}-\frac{\nabla^{\mu}\sqrt{\rho}}{\sqrt{\rho}}\Bigr). \label{daSdaSexpand}
\end{eqnarray}
Expanding this expression and substituting (\ref{cnt_idnty1}) into its right-hand side results in a somewhat more familiar expression
\begin{eqnarray}
\nabla_{\mu}S\nabla^{\mu}S=-\hbar^2\Biggl(\frac{\nabla_{\mu}\nabla^{\mu}\varphi}{\varphi}-\frac{\nabla_{\mu}\nabla^{\mu}\sqrt{\rho}}{\sqrt{\rho}}+\frac{i}{\hbar}\frac{\nabla_{\mu}\Omega^2}{\Omega^2}\nabla^{\mu}S\Biggr)=m^{2}\Omega^{2} \nonumber \label{EqmDaSDas}.
\end{eqnarray}
Here, we utilize (\ref{EoMsimp}) to retrieve the mass. By further rewriting $\nabla_{\mu}S=J_{\mu}/\rho$ in terms of the 4-vector current density $J_{\mu}$, multiplying both sides by the scalar fields $\varphi^{*}\varphi$, and reordering the resulting expression, we arrive at an equation of motion similar to that of Klein-Gordon
\begin{eqnarray}
\varphi^{*}\Box\varphi+\frac{i}{\hbar}\frac{\nabla_{\mu}\Omega^2}{\Omega^2}J^{\mu}+\frac{m^2}{\hbar^2}(\Omega^{2}-Q)\varphi^{*}\varphi=0 \label{WFeqnp1} 
\end{eqnarray}
Here, the 4-vector current density
$J^{\mu}=\frac{-i\hbar}{2}\Bigl(\varphi^{*}\nabla^{\mu}\varphi-\varphi\nabla^{\mu}\varphi^{*}\Bigr)=\rho\nabla^{\mu}S$ pertains to the non-conformal frame of reference (e.g. does not contain the conformal factor). In considering a linear-order constraint $\Omega^{2}=(1+Q)$, one arrives at the following equation of motion
\begin{eqnarray}
\varphi^{*}\Box\varphi+\frac{i}{\hbar}\frac{\nabla_{\mu}\Omega^2}{\Omega^2}J^{\mu}+\frac{m^2}{\hbar^2}\varphi^{*}\varphi=0. 
\label{WFeqn1} 
\end{eqnarray}
An exponential constraint would result in a more complicated pre-factor for the effective mass $(\Omega^{2}-Q)$. In expanding $\Omega^{2}=\textup{e}^{Q}$ into its Taylor series, it usually suffices to presume $Q^{2}\ll1$, aligning with the assumption of a linear-order constraint. 
For an exponential constraint, the ratio $\frac{\nabla_{\mu}\Omega^2}{\Omega^2}=\nabla_{\mu}\ln{\Omega^2}=\nabla_{\mu}Q$ is simply representative of the quantum force. 
In re-expressing (\ref{cnteqn}) into its current density form $\nabla_{\mu}J^{\mu}=-\frac{\nabla_{\mu}\Omega^2}{\Omega^2}J^{\mu}$, $J^{\mu}$ appears to provide a feedback to itself. 
It is therefore apparent that when the quantum force and current density $J^{\mu}$ are no longer orthogonal, current conservation breaks. This matter will become more clear in the next section. 
Using (\ref{cnteqn}), (\ref{WFeqn1}) can be re-expressed into its compact representation 
\begin{eqnarray}
\varphi^{*}\Box\varphi-\frac{i}{\hbar}\nabla_{\mu}J^{\mu}+\frac{m^2}{\hbar^2}\varphi^{*}\varphi=0. \label{QWFeqn1} 
\end{eqnarray}
One can further incline that the effective mass is analogous to the energy-momentum squared. As a result, a peculiar correspondence between the current density and effective mass can be deduced from the Hamilton-Jacobi equation
\begin{eqnarray}
\frac{1}{\hbar^2}\nabla_{\mu}S\nabla^{\mu}S=\frac{1}{\hbar^2}\frac{J_{\mu}J^{\mu}}{\rho^2}=\frac{1}{\hbar^2}m^2\Omega^{2}\;\;\to\;\;J_{\mu}J^{\mu}=m^2\Omega^{2}\rho^{2}. \label{current_cple}
\end{eqnarray}
Under the immature assumption that the \textit{effective} mass is a manifestation of the self-interaction of the current density, one can speculate that the particle may be interacting with a yet unexplored field variable. It is very much possible that there is another field variable which should accompany (\ref{WFeqn1}) whence the equations of motion are not conserved. After all, allowing the particle to arbitrarily dissipate energy, via the Noether current of (\ref{QWFeqn1}), would violate energy conservation. It is therefore highly compelling to define an alternative field variable which could mediate the particle's Noether current. 

Before introducing an alternative field variable, we further explore the implications of (\ref{WFeqnp1}). In utilizing the relation of (\ref{current_cple}), it is apparent that there exists a unique gauge connection associated to the relativistic quantum potential taking the form
\begin{eqnarray}
\varphi^{*}\mathcal{D}_{\mu}\mathcal{D}^{\mu}\varphi&=&\varphi^{*}\Bigl(\nabla_{\mu}-\frac{i}{\hbar}\nabla_{\mu}S\Bigr)\Bigl(\nabla^{\mu}-\frac{i}{\hbar}\nabla^{\mu}S\Bigr)\varphi \label{Qpsi} \\
&=&\varphi^{*}\Box\varphi-\frac{i}{\hbar}\nabla_{\mu}(\rho\nabla^{\mu}S)+\frac{1}{\hbar^{2}}\rho\nabla_{\mu}S\nabla^{\mu}S \nonumber \\
&=&\varphi^{*}\Box\varphi-\frac{i}{\hbar}\nabla_{\mu}J^{\mu}+\frac{m^{2}}{\hbar^{2}}\Omega^{2}\varphi^{*}\varphi \nonumber \\
&=&\frac{m^{2}}{\hbar^{2}}\varphi^{*}Q\varphi. \nonumber
\end{eqnarray}
The last line was attained by recognizing the third line's equivalence to the quantum potential, via (\ref{WFeqnp1}). 
In considering the field alignment scenario, the energy-momentum $\nabla_{\mu}S$ can be interpreted as being equivalent to a spatio-temporal derivative, resulting in a null gauge connection $\mathcal{D}_{\mu}\mathcal{D}^{\mu}\varphi\approx0$. This is analogous to the canonical representation of quantum mechanics, whereby the energy-momentum of the field variable exists only in phase space. In such a limit, the assumption of locality follows, along with the Poincar$\acute{\textup{e}}$ invariance of U(1) symmetric fields in quantum field theory.

Inserting the gauge connection $\mathcal{D}_{\mu}$ within the particle's 4-vector current density can further shine some light at its meaning
\begin{eqnarray}
J_{\mu}^{\mathcal{D}} =-\frac{i\hbar}{2}(\varphi^{*}\mathcal{D}_{\mu}\varphi-\varphi\mathcal{\widetilde{D}}_{\mu}\varphi^{*}) =J_{\mu}-\rho\nabla_{\mu}S=0. \label{CE_v1}
\end{eqnarray}
Here, $\mathcal{\widetilde{D}_{\mu}}=(\nabla_{\mu}+\frac{i}{\hbar}\nabla_{\mu}S)$ is simply the conjugate of $\mathcal{D}_{\mu}=(\nabla_{\mu}-\frac{i}{\hbar}\nabla_{\mu}S)$. By the result of (\ref{CE_v1}), the current density $J_{\mu}^{\mathcal{D}}$ is always null. This intuitively makes sense: the gauge connection effectively removes the phase contribution from $\nabla_{\mu}\varphi$; further suggesting that the proposed gauge connection is analogous to placing the particle within its comoving reference frame. The physical essence of the comoving frame can be better understood by again considering the field alignment scenario, whereby the classical energy-momentum relation is satisfied $\nabla_{\mu}S\nabla^{\mu}S\approx m^{2}$. In such a limit, the equation of motion results in $\mathcal{D}_{\mu}\mathcal{D}^{\mu}\varphi\approx0$, suggesting that the particle is effectively within a null accelerating frame-of-reference. 
Beyond this scenario $\mathcal{D}_{\mu}\mathcal{D}^{\mu}\varphi\neq0$, the particle is not exactly within the described null accelerating frame, rather traversing about trajectories dictated by the altered energy-momentum relation of (\ref{EoMsimp}). 

With this equation of motion (\ref{WFeqnp1}) and its corresponding gauge connection, we can now demonstrate the convergent behavior of the exponential conformal factor. That is, taking the product of an infinite number of quantum potentials $Q$ should amount to nullifying the significance of higher-order contributions $Q^{n}$, deeming $e^Q$ finite. For a single quantum potential, we have
\begin{eqnarray}
\frac{m^2}{\hbar^2}\varphi^{*}\frac{Q}{\Omega^{2}}\varphi=\varphi^{*}D_{\mu}D^{\mu}\varphi=\frac{1}{\Omega^{2}}\Bigl(\varphi^{*}\Box\varphi-\frac{i}{\hbar}\nabla_{\mu}J^{\mu}\Bigr)+\frac{m^{2}}{\hbar^2}\varphi^{*}\varphi. \label{gauge_v1}
\end{eqnarray}
Here, the quantum potential is divided by the conformal factor $\Omega^{2}$ to satisfy the proof which follows. As before, the gauge connection $\varphi^{*}\mathcal{D}_{\mu}\mathcal{D}^{\mu}\varphi$ simply results in the third line of (\ref{Qpsi}). Applying $Q$ consecutively $n$ times allows us to retrieve the following expression
\begin{eqnarray}
\frac{m^2}{\hbar^2}\varphi^{*}\frac{Q^{n}}{\Omega^{2n}}\varphi=\mathcal{U}_{n}(Q)\Bigl[\varphi^{*}\Box\varphi-\frac{i}{\hbar}\nabla_{\mu}J^{\mu}\Bigr]+\frac{m^2}{\hbar^2}\varphi^{*}\varphi. \label{gauge_v2}
\end{eqnarray}
Here, the function $\mathcal{U}_{n}$ can be represented as 
\begin{eqnarray}
\mathcal{U}_{n}(Q)=\frac{1}{\Omega^{2}}\Bigl(\frac{Q^{n-1}}{\Omega^{2(n-1)}}+\frac{Q^{n-2}}{\Omega^{2(n-2)}}+\cdots+\frac{Q}{\Omega^{2}}+1\Bigr).
\end{eqnarray}
In the limit of $n\to\infty$ and for $\Omega^{2}\gg Q$, it is apparent that $\mathcal{U}_{n}$ takes the form 
\begin{eqnarray}
\lim_{n\to\infty}\mathcal{U}_{n}(Q)=\frac{1}{\Omega^{2}}\frac{1}{1-Q/\Omega^{2}}. \label{gauge_v3}
\end{eqnarray}
In then multiplying the denominator of the above expression
\begin{eqnarray}
\frac{m^2}{\hbar^2}&&\lim_{n\to\infty}\varphi^{*}\bigl(\Omega^{2}-Q\bigr)Q^{n}\varphi \nonumber \\ &&=-\lim_{n\to\infty}\Omega^{2n}\Biggl[\Bigl(\varphi^{*}\Box\varphi+\frac{i}{\hbar}\nabla_{\mu}J^{\mu}\Bigr)+\frac{m^2}{\hbar^2}\Bigl(\Omega^{2}-Q\Bigr)\varphi\Biggr]=0, \label{gauge_v4}
\end{eqnarray}
we arrive at the required convergence criterion 
\begin{eqnarray}
\lim_{n\to\infty}\bigl(\Omega^{2}-Q\bigr)Q^{n}\approx\Omega^{2}\lim_{n\to\infty}Q^{n}=0\;\;\;\implies\;\;\;\lim_{n\to\infty}Q^{n}=0, \label{gauge_v5}
\end{eqnarray}
where the quantum potential on the right-hand side of (\ref{gauge_v4}) is expanded as a gauge-connection and, as result, cancels with its neighboring equation of motion. Equation (\ref{gauge_v5}) suggests that the equation of motion insures that, as long as $\Omega^{2}\gg Q$, the Taylor expanded exponential conformal factor remains a convergent series, even with a non-conserved Noether current. 

We now return back to the matter of a non-conserved current. By further introducing the acclaimed mediating field $\lambda$, via the constraint of (\ref{Actioneq}), the problem of arbitrarily dissipating energy could be alleviated. 
More specifically, the dissipated energy within (\ref{QWFeqn1}) could be more rigorously dictated by the mediating field and in certain limits, result in a conserved Noether current. 
The Bohmian equation of motion, containing the alleged mediating field, takes the form
\begin{equation}
\nabla_{\mu}S \nabla^{\mu}S-m^2\Omega^2+\frac{\hbar^2\Upsilon_{0}^{-1}}{2 
	\Omega^2\sqrt{\rho}}\Bigl[\Box{\Bigl(\frac{\lambda}{\sqrt{\rho}}\Bigr)}-\lambda\frac{\Box\sqrt{\rho}}{\rho}\Bigr]=0. \label{EqMotion} 
\end{equation}
Here, the procedure for defining the scalar field equation is not so different from before. After inserting (\ref{EqmDaSDas}) into (\ref{EqMotion}) and substituting the continuity equation (\ref{cnteqn}), we obtain a scalar field equation along with the mediating field 
\begin{equation}
\varphi^{*}\Box\varphi-\frac{i}{\hbar}\nabla_{\mu}J^{\mu}+\frac{m^2}{\hbar^2}(\Omega^{2}-Q)\varphi^{*}\varphi=\frac{\Upsilon_{0}^{-1}\sqrt{\rho}}{2\Omega^2}\Bigl[\Box\Bigl({\frac{\lambda}{\sqrt{\rho}}}\Bigr)-\lambda\frac{\Box{\sqrt{\rho}}}{\sqrt\rho}\Bigr] \label{WFeqnVacuum}. 
\end{equation}
In utilizing the equation of motion for the mediating field (\ref{LambdaEqlin}) and assuming a linear-order constraint $\Omega^{2}\approx(1+Q)$, (\ref{WFeqnVacuum}) can be re-expressed into a more pleasant form
\begin{eqnarray}
\varphi^{*}\Box\varphi-\frac{i}{\hbar}\nabla_{\mu}J^{\mu}+\frac{m^2}{\hbar^2}\varphi^{*}\varphi=\frac{\Upsilon_{0}^{-1}}{2\Omega^2}\Bigl[\Box-\frac{2m^2}{\hbar^2}\Bigr]\lambda \label{WFeqnVacuumF}. 
\end{eqnarray}
Here, the source term on the right-hand side resembles a relativistic diffusion equation. Equations (\ref{WFeqnVacuumF}) and (\ref{LambdaEqlin}) (in the linear-order limit of $\Omega^{2}$) are second-order coupled PDEs. The field $\lambda$ mediates the scalar field in a manner which will be described in the next section. More specifically, in the limit of a conserved mediating field ($\lambda\to\rho$), the source and dissipation term diminish. 

A fundamental question of interest is whether the mediating field should inherit an effective phase or always be real valued. In what follows, we will find the latter to be true in the limit of the field alignment scenario. To prove this, (\ref{WFeqnVacuumF}) is studied for a mediating field containing either a positive $\lambda=\rho e^{iS/\hbar}=\sqrt{\rho}\varphi$ or negative $\lambda=\rho e^{-iS/\hbar}=\sqrt{\rho}\varphi^{*}$ phase. In such a scenario, the scalar field equation can be written purely in terms of scalar field $\varphi$ but, unlike the field alignment scenario, the source term in (\ref{WFeqnVacuumF}) no longer vanishes. If the source term is invariant to the phase, the only physically significant representation of the mediating field must be real. By first substituting $\lambda=\sqrt{\rho}\varphi$ into (\ref{WFeqnVacuumF}), the scalar field equation can be written as
\begin{eqnarray}
\varphi^{*}\mathcal{D}_{\mu}\mathcal{D}^{\mu}\varphi+ \frac{e^{iS/\hbar}}{2\Upsilon_{0}}\varphi^{*}\Bigl(\mathcal{D}_{\mu}\mathcal{D}^{\mu}-\Box\Bigr)\varphi =0.\label{EoMphase}
\end{eqnarray}
Here, we assume that, for an infinitesimally small phase $S$, the classical energy-momentum relations are fairly well satisfied ($\nabla_{\mu}S\nabla^{\mu}S\approx m^{2}\implies\Omega^{2}\to1$) and, as a result, the gauge connection of (\ref{Qpsi}) can be imposed. Re-expressed differently
\begin{eqnarray}
\varphi^{*}\Box\varphi-\frac{i}{\hbar}\nabla_{\mu}J^{\mu}+\frac{m^2}{\hbar^2}\varphi^{*}\varphi+\frac{e^{iS/\hbar}}{2\Upsilon_{0}}\Bigl(-\frac{i}{\hbar}\nabla_{\mu}J^{\mu}+\frac{m^2}{\hbar^2}\varphi^{*}\varphi\Bigr)=0, \label{EoMphase_v0a}
\end{eqnarray}
it appears that the bracketed component on the left-hand side of (\ref{EoMphase}) varies the dissipation and mass term in a manner proportional to $e^{iS/\hbar}/2\Upsilon_{0}$.
In alternatively applying a negative phase to the mediating field $\lambda=\sqrt{\rho}\varphi^{*}=\rho e^{-iS/\hbar}$, the resulting equation of motion is
\begin{eqnarray}
\varphi^{*}\mathcal{D}_{\mu}\mathcal{D}^{\mu}\varphi+ \frac{e^{-iS/\hbar}}{2\Upsilon_{0}}\varphi\Bigl(\mathcal{\widetilde{D}}_{\mu}\mathcal{\widetilde{D}}^{\mu}-\Box\Bigr)\varphi^{*} =0 \label{EoMphase_v1}. 
\end{eqnarray}
One can then use the relation 
\begin{eqnarray}
\varphi^{*}\mathcal{D}_{\mu}\mathcal{D}^{\mu}\varphi=\varphi^{*}(Q\varphi)=(Q\varphi^{*})\varphi=(\mathcal{\widetilde{D}}_{\mu}\mathcal{\widetilde{D}}^{\mu}\varphi^{*})\varphi, \label{QPsiRel_v0}
\end{eqnarray}
along with the fact that $\varphi^{*}\Box\varphi\approx\varphi\Box\varphi^{*}$ in the limit of the field alignment scenario (for a conserved Noether current), to rewrite (\ref{EoMphase_v1}) into the following form 
\begin{eqnarray}
\varphi^{*}\mathcal{D}_{\mu}\mathcal{D}^{\mu}\varphi+ \frac{e^{-iS/\hbar}}{2\Upsilon_{0}}\varphi^{*}\Bigl(\mathcal{D}_{\mu}\mathcal{D}^{\mu}-\Box\Bigr)\varphi =0. \label{EoMphase_v2} 
\end{eqnarray}
Equations (\ref{EoMphase}) and (\ref{EoMphase_v2}) are the same up to a phase dictated by the mediated field. Given the solution of $\varphi$ in the limit of the field alignment scenario, the only way for (\ref{EoMphase}) and (\ref{EoMphase_v2}) to be physically consistent, is by deeming the phase prescribed within $\lambda$ to be non-existent. 
In the section which follows, we will demonstrate that the equation of motion for the mediating field is nothing more than a diffusion equation.

\section{Diffusion Equation}
Diffusion is, by definition, the translation of matter from a more to less dense region of spacetime. In what follows, we shall demonstrate that the mediating field effectively obeys a diffusion equation. It will be shown that, by the shift theorem, a diffusion equation can be used to articulate the manner by which information conforms to spacetime. 

In the limit of a linear-order conformal factor $\Omega^{2}=(1+Q)$, the most commonly expressed relativistic diffusion equation arises. By substituting the conservation condition of (\ref{cnstrnt1}) into the $\lambda$ equation (\ref{LambdaEqlin}), one arrives at 
\begin{eqnarray}
\Bigl[\Box-\frac{m^2}{\hbar^2}\Bigr]\lambda=0. \label{LambdaEqF}
\end{eqnarray}
By (\ref{cnstrnt1}), such an expression can only be satisfied in the field alignment scenario (e.g. $\lambda\to\rho$). In such a limit, the scalar field equation simplifies to its more familiar Klein-Gordon form, whereby the source and dissipation terms cease to exist. The non-relativistic form of the relativistic diffusion equation (\ref{LambdaEqF}) can be attained by applying the following non-relativistic approximation $\lambda\to\nu\textup{e}^{-\frac{m}{\hbar}t}$. 
Although (\ref{LambdaEqF}) may seem compelling, it is incomplete. Generally speaking, such relativistic diffusion equations can also lead to unstable solutions. This can be easily identified by applying the alternative non-relativistic approximation $\lambda\to\nu\textup{e}^{\frac{m}{\hbar}t}$. One possible way of overcoming this is by instead adopting the exponential conformal factor $\Omega^{2}=\textup{e}^{Q}$. In utilizing (\ref{LambdaEq}) along with the same conservation condition (\ref{cnstrnt1}), we arrive at
\begin{eqnarray}
\Bigl[\Box+\frac{1}{4}\Bigl(\frac{\nabla_{\mu}\lambda}{\lambda}\Bigr)\nabla^{\mu}-\frac{m^2}{\hbar^2}\Bigr]\lambda=0. \label{diffv0}
\end{eqnarray}
An expression for (\ref{diffv0}) beyond the field alignment scenario can be found by re-ordering (\ref{LambdaEq})
\begin{eqnarray}
\Bigl[\Box+\frac{1}{2}\Bigl(\frac{\nabla_{\mu}\lambda}{\lambda}-\frac{1}{2}\frac{\nabla_{\mu}\rho}{\rho}\Bigr)\nabla^{\mu}-\frac{m^{2}}{\hbar^{2}}\Bigr]\rho=0, \label{diffv1}
\end{eqnarray}
In utilizing the shift theorem, the terms in the round brackets of (\ref{diffv0}) and (\ref{diffv1}) can be described as the density's effective `shift.' The `shift' could be anything from a translation in spacetime to an arbitrary coordinate transformation. This concept will be further explored in the scalar field representation of the above diffusion equation. 
Representing (\ref{diffv0}) and (\ref{diffv1}) in their scalar field form results in a more interpretable coupled equations of motion. We begin by re-expressing (\ref{diffv1})
\begin{eqnarray}
2\frac{\Box\sqrt{\rho}}{\sqrt{\rho}}+\Bigl(\frac{\nabla_{\mu}\lambda}{\lambda}-\frac{\nabla_{\mu}\sqrt{\rho}}{\sqrt{\rho}}\Bigr)\frac{\nabla^{\mu}\sqrt{\rho}}{\sqrt{\rho}}=\frac{m^{2}}{\hbar^{2}}, \label{diffv2}
\end{eqnarray}
and utilizing the identities $\mathcal{D}_{\mu}\mathcal{D}^{\mu}\varphi=\frac{m^{2}}{\hbar^{2}}Q\varphi$ and $\mathcal{D}_{\mu}\varphi=\widetilde{\mathcal{D}}_{\mu}\varphi^{*}=\nabla_{\mu}\sqrt{\rho}$, to arrive at a scalar field-dependent equation of motion
\begin{eqnarray}
2\mathcal{D}_{\mu}\mathcal{D}^{\mu}\varphi+\Bigl(\frac{\nabla_{\mu}\lambda}{\lambda}-\frac{\widetilde{\mathcal{D}}_{\mu}\varphi^{*}}{\varphi^{*}}\Bigr)\mathcal{D}^{\mu}\varphi=\frac{m^2}{\hbar^2}\varphi. \label{diffv3}
\end{eqnarray}
In its field alignment form, the above expression can be further simplified
\begin{eqnarray}
2\varphi^{*}\mathcal{D}_{\mu}\mathcal{D}^{\mu}\varphi+\mathcal{\widetilde{D}}_{\mu}\varphi^{*}\mathcal{D}^{\mu}\varphi=\frac{m^2}{\hbar^2}\varphi^{*}\varphi. \label{diffv4}
\end{eqnarray}
An identity can then be utilized
\begin{eqnarray}
\mathcal{\widetilde{D}}_{\mu}\varphi^{*}\mathcal{D}^{\mu}\varphi&=&\Bigl(\nabla_{\mu}+\frac{i}{\hbar}\nabla_{\mu}S\Bigr)\varphi^{*}\Bigl(\nabla^{\mu}-\frac{i}{\hbar}\nabla^{\mu}S\Bigr)\varphi \nonumber \\
&=&\nabla_{\mu}\varphi^{*}\nabla^{\mu}\varphi+\frac{1}{\hbar^2}\varphi^{*}\varphi\nabla_{\mu}S\nabla^{\mu}S \nonumber \\
&=&\nabla_{\mu}\varphi^{*}\nabla^{\mu}\varphi+\Omega^{2}\frac{m^2}{\hbar^2}\varphi^{*}\varphi, \label{diffv5}
\end{eqnarray}
to result in the following expression
\begin{eqnarray}
2\varphi^{*}\mathcal{D}_{\mu}\mathcal{D}^{\mu}\varphi+\nabla_{\mu}\varphi^{*}\nabla^{\mu}\varphi=\varepsilon\frac{m^{2}}{\hbar^{2}}\varphi^{*}\varphi. \label{diffv6}
\end{eqnarray}
Here, an infinitesimally small value $\varepsilon$ is assigned to the mass as a result of an approximate field alignment scenario, whereby $(\Omega^{2}-1)m^{2}\approx(1+\varepsilon-1)m^{2}=\varepsilon{m}^{2}$. That is, the conservation conditions acquired from varying the Lagrangian are approximately satisfied, but not entirely, such that the results which follow hold in the \textit{limit} of $\lambda\to\rho$. 
In such a scenario, the effective mass $\Omega^{2}m^{2}$ converges to the usual mass $m^{2}$ and, as a result, the corresponding gauge connection of (\ref{Qpsi}) nullifies. 

In what follows, the shift theorem will be used to merge the coupled equations of motion into a single infinite-order PDE. 
To begin with, the coupled equations of motion take the form
\begin{eqnarray}
\mathcal{D}_{\mu}\mathcal{D}^{\mu}\varphi=0 \label{WFeqnVacuumF1}. 
\end{eqnarray}
\begin{eqnarray}
2\mathcal{D}_{\mu}\mathcal{D}^{\mu}\varphi+\frac{\nabla_{\mu}\varphi^{*}}{\varphi^{*}}\nabla^{\mu}\varphi=\varepsilon\frac{m^{2}}{\hbar^{2}}\varphi. \label{lambdaPsi_v3}
\end{eqnarray}
Here, the wave-like equation (\ref{WFeqnVacuumF1}) is assumed to contain some infinitesimally small dissipation term, one which will play a prior role in defining the corresponding single-particle Green's function. 
For perfectly aligned scalar and mediating fields, the simplest solution to the coupled equations of motion are attained by substituting (\ref{WFeqnVacuumF1}) into (\ref{lambdaPsi_v3}). 
In applying integration by parts and nullifying the boundary term, one results in a massless bosonic particle $\Box\varphi=0$. This massless expression naturally inherits a local conformal invariance.
Infinitesimally away from the field alignment scenario, a finite mass contribution $\varepsilon$ remains and an alternative procedure for merging the equations of motion can instead be adopted. 

One such method is in the use of the shift theorem. The shift theorem essentially allows for the information pertaining to the scalar field to be projected from a local domain of interest to some possibly non(quasi)-local region of space-time by virtue of a coordinate transformation. 
Such shift theorems have been studied in different contexts, such as p-adic string theory \cite{Padic_MathPhys,Calcagni2010,RollingTachyon}, where they were used to localize a nonlocal exponential operator by shifting the internal coordinates of its corresponding scalar field. 
In its simplest form, the shift theorem suggests that, for a uniform vector field $\gamma_{\mu}$, a projection operator $\textup{e}^{-\gamma^{\mu}\nabla_{\mu}}$ results in a translation of coordinates 
\begin{eqnarray}
\varphi(x'^{\alpha})=\textup{e}^{-\gamma^{\mu}\nabla_{\mu}}\varphi(x^{\alpha})=\int\widetilde{\varphi}(k^{\alpha})\textup{e}^{-i\gamma^{\mu}k_{\mu}}\textup{e}^{ix^{\mu}k_{\mu}}{d}^{4}k=\varphi(x^{\alpha}-\gamma^{\alpha}), \label{shiftTheorem}
\end{eqnarray}
Here, $x'^{\alpha}$ represents the transformed coordinate space, $k^{\alpha}$ is the spectral 4-vector, and $\widetilde{\varphi}(k^{\alpha})$ is the Fourier transformed scalar field. In this special scenario, the vector field $\gamma^{\mu}$ translates the internal coordinates $x^{\alpha}$ of the scalar field to an alternate coordinate representation $x'^{\alpha}=x^{\alpha}-\gamma^{\alpha}$, suggesting that the information of $\varphi$ at some local regime $x^{\alpha}$ may have some non-local correspondence. 
When $\gamma^{\mu}$ is no longer a uniform vector field, the shift operation of (\ref{shiftTheorem}) may result in a totally different coordinate transformation. For arbitrary $\gamma^{\mu}$, the projection operator $\textup{e}^{-\gamma^{\mu}(x)\nabla_{\mu}}$ results in a general coordinate transformation $\hat{\mathcal{O}}_{\;\;\beta}^{\alpha}$
\begin{eqnarray}
\varphi(x'^{\alpha})=\textup{e}^{-\gamma^{\mu}(x)\nabla_{\mu}}\varphi(x^{\alpha})=\varphi(\hat{\mathcal{O}}_{\;\;\beta}^{\alpha}x^{\beta}), \label{GenOperTheorem}
\end{eqnarray}
In this context, there are an infinite number of possible coordinate transformation which could effectively shift the scalar field. As an alternative example, for a vector field $\gamma^{\mu}(x)=ax^{\mu}$, the operator $\hat{\mathcal{O}}_{\;\;\beta}^{\alpha}$ scales the internal coordinates of the scalar field ($\hat{\mathcal{O}}_{\;\;\beta}^{\alpha}x^{\beta}\to \textup{e}^{a}x^{\alpha}$). 

Of all the possible ways of merging (\ref{WFeqnVacuumF1}) and (\ref{lambdaPsi_v3}) into a single equation of motion, doing so with the shift theorem provides the intuition into the underlying mechanism of the relativistic diffusion equation. 
In what follows, the shift theorem will be used to transform the internal coordinates of the scalar field contained within the wave-like equation. 
Using (\ref{lambdaPsi_v3}), the transformation operator $\hat{\mathcal{O}}_{\;\;\beta}^{\alpha}$ can then be interpreted as containing the geometrical properties inherited by the now \textit{non-local} scalar field. To demonstrate this, (\ref{lambdaPsi_v3}) can first be rewritten as
\begin{eqnarray}
\left(\frac{2h^2}{m^2}\mathcal{D}_{\mu}\mathcal{D}^{\mu}-\varepsilon\right)\varphi=-\gamma^{\mu}(x)\nabla_{\mu}\varphi, \label{lambdaPsi_v4}
\end{eqnarray}
where $\gamma^{\mu}(x)=\frac{h^2}{m^2}\frac{\nabla_{\mu}\varphi^{*}}{\varphi^{*}}$. The projection operator can then be used to impose a transformation on the internal coordinates of the scalar field $\varphi$
\begin{eqnarray}
\varphi(x'^{\alpha})=\varphi(\hat{\mathcal{O}}_{\;\;\beta}^{\alpha}x^{\beta})=\textup{e}^{-\gamma^{\mu}\nabla_{\mu}}\varphi(x^{\alpha})=\textup{e}^{\Bigl(\frac{2h^2}{m^2}\mathcal{D}_{\mu}\mathcal{D}^{\mu}-\varepsilon\Bigr)}\varphi(x^{\alpha}). \label{lambdaPsi_v5}
\end{eqnarray}
In what follows, we assume $\textup{e}^{\varepsilon}\approx1$. Applying this projection operator onto the scalar field equation, (\ref{WFeqnVacuumF1}) can be re-written as
\begin{eqnarray}
\mathcal{D}_{\mu}\mathcal{D}^{\mu}\varphi(x'^{\alpha})=\mathcal{D}_{\mu}\mathcal{D}^{\mu}\Bigl(\textup{e}^{\frac{2h^2}{m^2}\mathcal{D}_{\nu}\mathcal{D}^{\nu}}\varphi(x^{\alpha})\Bigr)=0 \label{WFeqnVacuumF2}. 
\end{eqnarray}
Resulting in an infinite-order PDE
\begin{eqnarray}
\textup{e}^{\frac{2h^2}{m^2}\mathcal{D}_{\nu}\mathcal{D}^{\nu}}\mathcal{D}_{\mu}\mathcal{D}^{\mu}\varphi=0 \label{WFeqnVacuumF3}. 
\end{eqnarray}
In the typical quantum theory, such diffusion behavior is not accounted for. It can then be hypothesized that, unlike the usual quantum theory, occurring in the perfect field alignment scenario, the local attribute of the scalar field no longer holds for infinitesimally misaligned scalar and mediating fields. 
In this sense, the diffusion equation of the proposed theory alters the standard form of quantum mechanics.

This can be exemplified canonically by studying the implication of (\ref{WFeqnVacuumF3}) on the standard form of the commutation relations.
To simplify matters, we begin with the commutation relations resulting from a linear-order conformal factor. 
Given a metric tensor $g_{\mu\nu}$ with signature $(+,-,-,-)$, the commutation relations can be derived heuristically using the standard 4-vector momentum $\hat{p}$ and coordinate $\hat{x}$ operators
\begin{eqnarray}
[\hat{x}_{\mu},\hat{p}_{\nu}]=i\hbar g_{\mu\nu}.
\end{eqnarray}
In the flat-space scenario $g_{\mu\nu}\to\eta_{\mu\nu}$, this relation is not altered, resulting in the usual quantum mechanics. For a linear-order conformal factor, the commutation relations in a flat-space background ${g}_{\mu\nu}\to\Omega^2 \eta_{\mu\nu}$ are represented by some \textit{generalized} momentum operator $\hat{P}_\nu$
\begin{eqnarray}
[\hat{x}_{\mu},\hat{P}_{\nu}]=i\hbar g_{\mu\nu}=i\hbar \Omega^2 \eta_{\mu\nu}=i\hbar (1+Q) \eta_{\mu\nu}. \label{comRel0}
\end{eqnarray}
By further utilizing the gauge connection of (\ref{Qpsi}), (\ref{comRel0}) can be re-expressed into an operator form 
\begin{eqnarray}
[\hat{x}_{\mu},\hat{P}_{\nu}]\varphi=i\hbar \eta_{\mu\nu} \Bigl(1+\frac{\hbar^2}{m^2}\mathcal{D}_\alpha \mathcal{D}^\alpha \Bigr)\varphi.
\end{eqnarray}
Here, the relativistic quantum potential is replaced by the gauge connection $Q\varphi=\frac{\hbar^{2}}{m^{2}}\mathcal{D}_\alpha \mathcal{D}^\alpha\varphi$ and a scalar field $\varphi$ remains outside of the brackets on the right- and left-hand sides. In the field alignment scenario, the resulting gauge-connection can be canonically approximated by the usual momentum operator $\hbar^{2}\mathcal{D}_{\alpha}\mathcal{D}^{\alpha}\approx-\frac{\hat{p}_\alpha\hat{p}^\alpha}{2}$ via (\ref{lambdaPsi_v3})
\begin{eqnarray}
[\hat{x}_{\mu},\hat{P}_{\nu}]=i\hbar \eta_{\mu\nu} \Bigl(1-\frac{1}{m^2}\frac{\hat{p}_{\alpha}\hat{p}^{\alpha}}{2}\Bigr),
\end{eqnarray}
And the generalized momentum operator $\hat{P}_\mu$ can then be expressed as a function of $\hat{p}_\mu$
\begin{eqnarray}
\hat{P}_{\mu}=\Bigl(1-\frac{1}{m^{2}}\frac{\hat{p}_{\alpha}\hat{p}^{\alpha}}{2}\Bigr)\hat{p}_{\mu}. \label{CRfirst}
\end{eqnarray}
Higher order contributions of the momentum operator can be obtained by considering the exponential extension of (\ref{comRel0}). By considering the infinite-order PDE of the scalar field in (\ref{WFeqnVacuumF3}) and recognizing the fact that the contracted covariant derivative $\mathcal{D}_\alpha \mathcal{D}^\alpha$ commutes with itself, the exponential operator $e^{\frac{\hbar^2}{m^2} \mathcal{D}_\alpha \mathcal{D}^\alpha}$ can instead be adopted
\begin{eqnarray}
[\hat{x}_{\mu},\hat{P}_{\nu}]=i\hbar\Omega^2 \eta_{\mu\nu}=i\hbar \eta_{\mu\nu} e^{\frac{\hbar^2}{m^2}\mathcal{D}_\alpha \mathcal{D}^\alpha} \label{ComRel0}
\end{eqnarray}
\begin{eqnarray}
[\hat{x}_{\mu},\hat{P}_{\nu}]=i\hbar \eta_{\mu\nu} \Biggl\{1-\frac{1}{m^2}\frac{\hat{p}_{\alpha}\hat{p}^{\alpha}}{2}+\frac{1}{2!}\frac{1}{m^4}\Bigl(\frac{\hat{p}_{\alpha}\hat{p}^{\alpha}}{2}\Bigr)^{2}-\frac{1}{3!}\frac{1}{m^6}\Bigl(\frac{\hat{p}_{\alpha}\hat{p}^{\alpha}}{2}\Bigr)^{3}+\cdots\Biggr\}. \label{ComRel1}
\end{eqnarray}
Here, the exponential operator has been expanded into its Taylor series and, as before, $\hbar^{2}\mathcal{D}_{\alpha}\mathcal{D}^{\alpha}\to-\frac{\hat{p}_{\alpha}\hat{p}^{\alpha}}{2}$. Although the above commutation relation is heuristically defined via canonical operators, it clearly reproduces the infinite-order scalar field equation (\ref{WFeqnVacuumF3}). The corresponding generalized momentum operator takes the form
\begin{eqnarray}
\hat{P}_\mu =\Biggl\{1 - \frac{1}{m^2}\frac{\hat{p}_{\alpha}\hat{p}^{\alpha}}{2} + \frac{1}{2!m^4}\Bigl(\frac{\hat{p}_{\alpha}\hat{p}^{\alpha}}{2}\Bigr)^2 - ... 
\Biggr\}\hat{p}_\mu. \label{GenPmu}
\end{eqnarray}
The higher-order contributions of the generalized momentum operator clearly become more important for particles of smaller mass and higher energies. 
Such corrections would therefore not be necessary for particles of lower energy and higher mass. 
One possible physical interpretation of such commutation relations is that the higher-order terms can be perceived as a consequence of matter's self-interaction. Such self-interactions, to first-order, have been shown to account for vacuum energy corrections in attaining the Lamb shift~\cite{SidharthLambRevw}. One can perceive this as a kind-of feedback mechanism with a background field, one which, in this particular scenario, must conform with itself. This notion is well established in the dealing with Einstein's equation, whereby the stress-energy constantly seeks to balance itself with its associated metric tensor in characterizing its encompassing manifold structure. 

The commutation relations of (\ref{ComRel1}) can further be used to articulate a generalized uncertainty principle, via the Cauchy-Schwartz identity, in assuming a null expectation value of any single operator $\expval{\hat{A}}=0$
\begin{align}
\expval{\hat{x}\hat{x}}{\varphi}\expval{\hat{P}\hat{P}}{\varphi}&\geq\abs{\expval{\hat{x}\hat{P}}{\varphi}}^{2} \label{UnctyPrinc_v0} \\
(\Delta x)^{2}(\Delta P)^{2}&\geq \abs{\textrm{Im}{\expval{\hat{x}\hat{P}}{\varphi}}}^{2} \label{UnctyPrinc_v1} \\
(\Delta x)^{2}(\Delta P)^{2}&\geq\abs{\frac{1}{2i}\bigl[\expval{\hat{x}\hat{P}}{\varphi}-\expval{\hat{x}\hat{P}}{\varphi}^{*}\bigr]}^{2}. \label{UnctyPrinc_v2} 
\end{align}
Here, (\ref{UnctyPrinc_v1}) is attained in presuming that $\abs{\expval{\hat{x}\hat{P}}{\varphi}}\geq\abs{\textrm{Im}{\expval{\hat{x}\hat{P}}{\varphi}}}$. 
For the generalized momentum operator $\hat{P}$ of (\ref{ComRel1}), it further suffices to assume that $\expval{\hat{x}\hat{P}}{\varphi}^{*}=\expval{\hat{P}\hat{x}}{\varphi}$. 
In adopting the notation $\expval{\hat{x}\hat{P}}{\varphi}=\expval{\hat{x}\hat{P}}$, we then arrive at the following expression
\begin{align}
(\Delta{x})^{2}(\Delta{P})^{2}&\geq-\frac{1}{4}\abs{\expval{\bigl[\hat{x},\hat{P}\bigr]}}^{2}. \label{UnctyPrinc_v4}
\end{align}
The modified uncertainty principle can then be expressed as 
\begin{eqnarray}
&(\Delta{x})(\Delta{P})\geq\frac{\hbar}{2}\abs{1-\expval{\frac{1}{2m^{2}}\hat{p}^{2}-\frac{1}{8m^{4}}\hat{p}^{4}+\cdots}} \label{UnctyPrinc_fin} \\
&(\Delta{x})(\Delta{P})\geq\frac{\hbar}{2}\abs{1-\mathcal{E}}. \nonumber
\end{eqnarray}
Here, $\mathcal{E}$ is simply the expectation value of the higher-order energy-momentum contributions. For practical purposes, the covariant and contravariant indices of the 4-vector momentum operator $\hat{p}$ are appended for the time being. At relatively low energies, Heisenberg's uncertainty principle is retrieved. At higher energies, the result indicates that the aforementioned feedback contributions play a prior role. To make more sense of (\ref{UnctyPrinc_fin}), it is alternatively regrouped into its exponential form
\begin{align}
(\Delta{x})(\Delta{P})&\geq\frac{\hbar}{2}\abs{\expval{\textup{e}^{\frac{\hbar^2}{m^2}\mathcal{D}_{\mu}\mathcal{D}^{\mu}}}} \label{UnctyPrinc_finS1} \\
(\Delta{x})(\Delta{P})&\geq\frac{\hbar}{2}\abs{\int_{-\infty}^{\infty}(\varphi^{*}\textup{e}^{\frac{\hbar^2}{m^2}\mathcal{D}_{\mu}\mathcal{D}^{\mu}}\varphi) dx}. \label{UnctyPrinc_finS2}
\end{align}
As before, the shift theorem can be utilized to rewrite the operator in accordance with the diffusion equation (\ref{lambdaPsi_v4}).
Ignoring the infinitesimal contribution $\varepsilon$ and in the limit of the field alignment scenario, we arrive at
\begin{eqnarray}
\frac{\hbar^2}{m^2}\varphi^{*}\mathcal{D}_{\mu}\mathcal{D}^{\mu}\varphi=-\frac{1}{2}\frac{\hbar^2}{m^2}\nabla^{\mu}\varphi^{*}\nabla_{\mu}\varphi \label{StatLambda0} \\
\frac{\hbar^2}{m^2}\mathcal{D}_{\mu}\mathcal{D}^{\mu}\varphi=-\gamma^{\mu}\nabla_{\mu}\varphi. \label{StatLambda1}
\end{eqnarray}
In setting $\gamma^{\mu}=\frac{1}{2}\frac{\hbar^2}{m^2}\frac{\nabla^{\mu}\varphi^{*}}{\varphi^{*}}$, the exponential operator in (\ref{UnctyPrinc_finS2}) can be re-expressed as
\begin{eqnarray}
e^{\frac{\hbar^2}{m^2}\mathcal{D}_{\mu}\mathcal{D}^{\mu}}\varphi(x)=e^{-\gamma^{\mu}\nabla_{\mu}}\varphi(x)=\varphi(\hat{\mathcal{O}}x). \label{StatLambda2}
\end{eqnarray}
Here, and as before, $\hat{\mathcal{O}}$ is the operator acting on $x$ for some given vector field $\gamma^{\mu}$. The generalized uncertainty principle then takes the form
\begin{eqnarray}
(\Delta{x})(\Delta{P})\geq\frac{\hbar}{2}\abs{\expval{\textup{e}^{-\gamma^{\mu}\nabla_{\mu}}}} \label{UnctyPrinc_finS3} 
\end{eqnarray}
\begin{eqnarray}
(\Delta{x})(\Delta{P})\geq\frac{\hbar}{2}\abs{\int_{-\infty}^{\infty}\bigl[\varphi^{*}(x)\varphi(\hat{\mathcal{O}}x)\bigr] dx}. \label{UnctyPrinc_finS4}
\end{eqnarray}
When the coordinate system is invariant to the operator $\hat{\mathcal{O}}x\approx{x}$, the generalized uncertainty principle reduces to its original form $\abs{\int_{-\infty}^{\infty}\bigl[\varphi^{*}(x)\varphi(\hat{\mathcal{O}}x)\bigr] dx}=1$. The form of the operator $\hat{\mathcal{O}}$ that imposes such invariance would have to be that of the identity operator. 
Otherwise, for some nontrivial coordinate transformation $\hat{\mathcal{O}}x\ne{x}$, the lower bound in uncertainty can converge to the classical limit, whereby $\abs{\int_{-\infty}^{\infty}\bigl[\varphi^{*}(x)\varphi(\hat{\mathcal{O}}x)\bigr]dx}\approx0$. Furthermore, it is clear that, given a positive-definite norm of the integral, $\mathcal{E}$ in (\ref{UnctyPrinc_fin}) must be positive-definite to cohere with (\ref{UnctyPrinc_finS4}). 

The feedback mechanism encountered in the commutation relations and generalized uncertainty principle hinders the question of whether such corrections can be articulated by a manifold structure, rather than canonical operators. 
In what follows, we conjecture a representation theorem which implies that, rather than a generalized momentum operator, one can equivalently resolve the corrections to the uncertainty in phase space using differential forms. 
As a result, in the argument which follows, it is the length scale, rather than the phase, which articulates the corrections to the quantum mechanical uncertainty. 
To better grasp this, we further explore differential forms and their corresponding operators within a geometrical framework. 
By analyzing the infinitesimal length element $ds^2$ and fixing the momentum operator to its classical form, one can see that the spacetime naturally allows for deformations which become more significant for smaller masses in smaller length scales
\begin{eqnarray}
ds^2=dX_{\mu}dX^{\mu}=\Omega^2 \eta_{\mu\nu}dx^{\mu}dx^{\nu}
\end{eqnarray}
\begin{eqnarray}
dX_{\mu}dX^{\mu}\sqrt{\rho}=dx_{\mu}dx^{\mu}\Bigl(1+\frac{\hbar^2}{m^2}\nabla_{\alpha}\nabla^{\alpha}\Bigr)\sqrt{\rho}.
\end{eqnarray}
Here, ${dX}^{\mu}$ is the differential form and $\sqrt{\rho}$ appears as a consequence of the relativistic quantum potential. By further taking the square root of both sides and adopting the approximation 
$\Bigl(1+\frac{\hbar^2}{m^2}\nabla_{\alpha}\nabla^{\alpha}\Bigr)^{\frac{1}{2}}\approx 
\Bigl(1+\frac{\hbar^2}{2m^2}\nabla_{\alpha}\nabla^{\alpha}\Bigr)$,
\begin{eqnarray}
d\hat{X}^{\mu}=dx^{\mu}\Bigl(1+\frac{\hbar^2}{2m^2}\nabla_{\alpha}\nabla^{\alpha}\Bigl), \label{difFormOpr0}
\end{eqnarray} 
we arrive at the linear-order representation of what we refer to as the \textit{differential form operator} $d\hat{X}^{\mu}$. 
The exponential conformal factor $\Omega^2=e^{Q}$ can be considered in a similar fashion to (\ref{GenPmu}) with a key difference: the differential form operator $d\hat{X}^{\mu}$ must be confined to second order to satisfy the bilinear form of the metric, and therefore must act on the density $d\hat{X}^{\mu}\sqrt{\rho}=dX^{\mu}\sqrt{\rho}$,
\begin{eqnarray}
dX_{\mu}dX^{\mu}= \eta_{\mu\nu}{dx}^{\mu}{dx}^{\nu}\Bigl[1+Q+\frac{1}{2!}Q^{2}+\frac{1}{3!}Q^{3}+\cdots\Bigr]. \label{difform_v0}
\end{eqnarray}
Here, the corresponding momentum $p^\mu$ takes on the classical form $p^\mu=m\frac{d\hat{X}^{\mu}}{d\tau}$, where $\tau$ is simply the affine parameter associated to the metric tensor. 
In their spatio-temporal representation, $d\hat{X}^{\mu}$ and $p^\mu$ are both real quantities. They are treated as two separate entities, such that $d\hat{X}^{\mu}$ defines the differential manifold and $p^\mu$ articulates the geodesic equation.

My hope is that in considering the aforementioned feedback, one originating purely from a quantum mechanical origin, accelerative contributions, analogous to those considered within general relativity, can be can be properly characterized within a generalized theory of quantum mechanics. 
Whether gravitational phenomena result from such a coupling is not at all clear from the work done thus far. 
Regardless, I conjecture that one can define a representation theorem for the incorporation of such accelerative contributions by modifying the quantum mechanical commutation relations in one of two ways:
\begin{eqnarray}
(\hat{x}_\mu,\hat{P}_\nu)\iff(d\hat{X}_\mu,p_\nu).
\end{eqnarray}
Such a representation theorem suggests that one can equivalently adopt the modified momentum operator in a flat-spacetime or the differential form in a curved-spacetime structure. 
Here, a `flat-spacetime' does not imply that gravity cannot be incorporated as an affine-connection in the covariant derivatives of the modified scalar field equation, it is simply meant to indicate the manner by which the corrections to the uncertainty are characterized.
In the section which follows, we derive the propagator within the flat-space representation $(\hat{x}_\mu,\hat{P}_\nu)$ to elucidate the nature of the modified scalar field equation.

\section{Propagator}
In what follows, we define the single-particle propagator corresponding to the previously derived infinite-order scalar field equation. 
Using the exponential form of the momentum operator in (\ref{GenPmu}), (\ref{WFeqnVacuumF3}) can be replicated 
\begin{eqnarray}
\hat{P}_\mu\hat{P}^\mu \varphi =0 \implies \hbar^2\,e^{\frac{2\hbar^2}{m^2}\mathcal{D}_{\mu}\mathcal{D}^{\mu}}\,\mathcal{D}_{\mu}\mathcal{D}^{\mu}\varphi=0, \label{InfOrderEoM}
\end{eqnarray}
and the corresponding propagator can be determined under the assumption of an approximately local scalar field
\begin{eqnarray}
\Bigl(\hbar^2\,e^{\frac{2\hbar^2}{m^2}\mathcal{D}_{\mu}\mathcal{D}^{\mu}}\,\mathcal{D}_{\mu}\mathcal{D}^{\mu}\Bigr) G(x,x')=-i\delta^{4}(x-x'),
\end{eqnarray}
The factor of two in (\ref{InfOrderEoM}) can only be attained in adopting an exponential conformal factor and was critical in attaining the corresponding infinite-order PDE. The assumption of locality, via the Dirac-Delta function $\delta^{4}(x)$, holds only in the limit of the field alignment scenario. The spectral representation of the propagator $\tilde{G}(k)$ can further be articulated as
\begin{eqnarray}
\tilde{G}(k)=\frac{i\,\exp{\Bigr(-\frac{2}{m^2}(-\hbar^2 k^2+m^2-i\hbar\tilde{\epsilon}(k))\Bigl)}}{(- m^2 + i\hbar\tilde{\epsilon}(k)+\hbar^2 k^2)}.
\end{eqnarray}
Here,  ${\epsilon}(k)=\frac{1}{(2\pi)^4}\int{\Bigl(\frac{\nabla_{\mu}J^{\mu}}{\rho}\Bigr)e^{ik_{\mu}x^{\mu}}d^4x}$ and $\tilde{\epsilon}(k)=[{\epsilon}(k)*\tilde{G}(k)]\tilde{G}^{-1}(k)$. $\tilde{\epsilon}$ is an imaginary contribution resulting from the gauge connection in (\ref{Qpsi}). $\epsilon(k)$ must convolve with $\tilde{G}(k)$ to properly characterize $\tilde{\epsilon}(k)$. The imaginary contribution $\tilde{\epsilon}$ within the Green's function seems to contain $\tilde{G}$, suggesting the Green's function can only be characterized self-consistently. Such an outcome is analogous to the aforementioned feedback mechanism encountered in heuristically attaining the generalized uncertainty principle. The function $\tilde{\epsilon}$, as infinitesimal as it may be, is unique in that it effectively shifts the poles away from the real axis as is done in attaining the retarded Feynman propagator. 
Furthermore, the proposed propagator eliminates the UV- and IR-Divergences appearing in Feynman diagrams.
In a non-renormalizable theory, the exponential in the denominator always inherits higher-order powers of $k$ to counter balance any terms appearing in the numerator for any loop order. 
Similarly, the imaginary component in the denominator will stay finite in the limit of $k\to 0$. 

Although the higher-order contributions appearing in the aforementioned commutation relations lead to seemingly interesting generalizations of the equation of motion, higher-orders can prove to be disadvantageous both conceptually and numerically. 
The main criticism concerned with the higher-order derivative theories is their associated Ostrogradsky instability. 
In string theories, non-local equations with infinitely many powers of the d'Alembertian operator are frequently studied~\cite{Freund87padic,Zwiebach_String2002,Gianluca2008}. 
One of their main issues is that the system Hamiltonian linearly depends on some of the momentum coordinates, allowing the momentum to freely take on negative values. Negative values of momentum lead to Hamiltonians unbounded from below, resulting in the infamous Ostrogradsky instability~\cite{OstroInstab2015}. 
Ostrogradsky instability theorem states that ``\textit{For any non-degenerate theory whose dynamical variable is higher than second-order in the time derivative, there exists a linear instability}''~\cite{Woodard2007Ostro,Woodard2015Ostro}. 
This creates issues like negative norm states or ghost states in the corresponding quantum theory. 
Mannheim and Bender proposed that PT-Symmetric Hamiltonians~\cite{Bender1,Bender2,Bender3,Bender4,Manheim_PTSym,BenderManheimGhost08} can effectively remove such ghosts from higher-order derivative theories. 
Others have also shown that one can exorcise the ghosts, and hence eliminate such instabilities, by using constraints to reduce the dimensionality of the phase-space~\cite{Chen2013ostro,BarnabyKamranOstro}. 

Still, a question arises: can the alleged instability associated to higher-order derivative theories be directly applied to infinite-order equations? Recently~\cite{BarnabyKamranOstro}, N. Barnaby and N. Kamran have indicated that equations with infinitely many derivatives can never consistently be viewed as the $N\to \infty$ limit of some $N$-th order equation. They have also shown that differential equations of infinite-order do not generically admit infinitely many initial data. There is a crucial difference between finite-order and infinite-order derivative theories; the former acts locally on the field variable while the latter nonlocally on the field variable. Therefore, one should be careful in rejecting infinite-order theories simply on the basis of Ostrogradsky instability.

\section{Summary \& Outlook}
In this manuscript, wave-like and diffusion equations of motion were articulated from a conformally transformed Einstein-Hilbert action composed of a Klein-Gordon field and a unique constraint. 
Higher-order contributions of the Taylor expanded exponential conformal factor were found to gradually nullify, deeming the conformal factor finite in the limit of the field alignment scenario. 
A scalar field equation, containing a source and dissipation term, was derived from its Bohmian interpretation by merging the corresponding continuity equation with the equation of motion for the density. 
The dissipation term took the form of the non-conserved Noether current ($\nabla_{\mu}J^{\mu}\ne0$). 
Both source and dissipation nullified in the limit of the field alignment scenario, whereby the conservation condition of (\ref{cnstrnt1}) was satisfied. 
The constraint's Lagrange multiplier $\lambda$ was shown to act as the mediator for the source and dissipation terms. 
Furthermore, the mediating field was dictated by a diffusion equation and was suggested to transform the internal coordinates of the scalar field, projecting the local information of the field to different regions of spacetime. 
In the field alignment scenario, the two coupled equations of motion were merged into an infinite-order PDE in utilizing the shift theorem. 
Modified canonical commutation relations and a corresponding generalized uncertainty principle were heuristically defined in the limit of the field alignment scenario. 
A representation theorem, conjecturing a one-to-one correspondence between momentum and differential form operators, was introduced.
Finally, the generalized uncertainty principle was conjectured to eliminate infrared and ultraviolet divergences within the infinite-order PDEs propagator. 

Much of the formal work done here only holds in the limit of the field alignment scenario, whereby the mediating and scalar fields approximately align. Further work needs to be done to understand the physical essence of matter's alleged self-interaction beyond the field alignment scenario. This would require incorporating the conservation conditions (\ref{cnstrnt1}) as boundary terms within the equation of motion. 

The non-local theory presented here can best be summarized by the coupled second-order equations of motion (\ref{WFeqnVacuumF1})-(\ref{lambdaPsi_v3}). With both a scalar and mediating field, the theory appears to acquire a rich collection of physical scenarios yet to be determined. 
Given the misalignment of the scalar and mediating fields is relatively small, the boundary terms can be ignored and the gauge connection can be approximately imposed, so as to result in the generalized coupled equations of motion
\begin{eqnarray}
\varphi^{*}\mathcal{D}_{\mu}\mathcal{D}^{\mu}\varphi=\Bigl[\frac{\Upsilon_{0}^{-1}}{2\Omega^{2}}\Bigl(\Box-\frac{2m^2}{\hbar^2}(1-Q)\Bigr)\lambda\Bigr] \label{WFeqnVacuumFExp} 
\end{eqnarray}
\begin{eqnarray}
2\mathcal{D}_{\mu}\mathcal{D}^{\mu}\varphi+\Bigl(\frac{\nabla_{\mu}\lambda}{\lambda}-\frac{\widetilde{\mathcal{D}}_{\mu}\varphi^{*}}{\varphi^{*}}\Bigr)\mathcal{D}^{\mu}\varphi=\frac{m^2}{\hbar^2}\varphi. \label{LambdaNiceEqExp}
\end{eqnarray}
It is my hope that these equations can serve as a path for acquiring the needed corrections to quantum mechanics at smaller length scales. 
The mediating field's diffusion-like properties and the scalar field's inherent self-interaction all hinder at the possibility that $\lambda$ may be strongly characteristic of an aether-like entity governing the non(quasi)-local configuration of matter within spacetime. 
Whether it directly corresponds to the quantum vacuum is yet to be determined.

\bibliographystyle{JHEP}
\providecommand{\href}[2]{#2}\begingroup\raggedright\endgroup

\end{document}